\documentclass[aps,preprint]{revtex4}

\usepackage{amssymb}
\usepackage{amsmath}
\usepackage{epsfig}
\usepackage{epstopdf}
\usepackage{bm}
\usepackage{graphicx,epsfig}
\usepackage{mathrsfs}
\usepackage{dcolumn}
\usepackage{color}
\usepackage{natbib}
\usepackage{CJK}
\def\be{\begin{equation}}
\def\ee{\end{equation}}
\def\bea{\begin{eqnarray}}
\def\eea{\end{eqnarray}}

\allowdisplaybreaks[2]

\begin{document}

\title{Spatiotemporal coupled-mode equations for arbitrary pulse transformation}
\author{Zhaohui Dong$^{1}$, Xianfeng Chen$^{1,2,3}$, and Luqi Yuan$^{1,*}$}
\affiliation{$^1$State Key Laboratory of Advanced Optical Communication Systems and Networks, School of Physics and Astronomy, Shanghai Jiao Tong University, Shanghai 200240, China \\
$^2$Shanghai Research Center for Quantum Sciences, Shanghai 201315, China \\
$^3$Collaborative Innovation Center of Light Manipulation and Applications, Shandong Normal University, Jinan 250358, China \\
$^{\ast} $yuanluqi@sjtu.edu.cn}

\begin{abstract} 
Spatiotemporal modulation offers a variety of opportunities for light manipulations. 
In this paper, we propose a way towards arbitrary transformation for pulses sequentially propagating within one waveguide in space via temporal waveguide coupling. 
The temporal waveguide coupling operation is achieved by spatiotemporally modulating the refractive index of the spatial waveguide with a traveling wave through segmented electrodes. 
We derive the temporal coupled-mode equations and discuss how systematic parameters affect the temporal coupling coefficients. 
We further demonstrated a temporal Mach-Zehnder interferometer and universal multiport interferometer, which enables arbitrary unitary transformation for pulses. 
We showcase a universal approach for transforming pulses among coupled temporal waveguides, which requires only one spatial waveguide under spatiotemporal modulation, and hence provide a flexible, compact, and highly compatible method for optical signal processing in time domain.
\end{abstract}

\maketitle

\section*{Introduction}
Time-varying media brings intriguing opportunities for wave manipulation in photonics \cite{engheta2021metamaterials, galiffi2022photonics, yin2022floquet, yuan2022temporal, pacheco2022time} and hence attracts growing interest in both physics community and optical engineering.
In particular, by combining both temporal and spatial degrees of freedom, the photonic systems undergoing spatiotemporal modulations recently emerge as new platforms for controlling light simultaneously in space and time \cite{shaltout2019spatiotemporal, mock2019magnet, deck2019uniform, tian2020hybrid, panuski2022full, gurses2022enhancing}. 
Utilizing this powerful approach, researchers explore many exotic phenomena which cannot be realized in a static medium, such as luminal amplification \cite{galiffi2019broadband, pendry2021gain, pendry2021gainin}, Fresnel drag \cite{huidobro2019fresnel, xu2022diffusive}, magnet-free nonreciprocal systems \cite{wang2018observation, sounas2017non, taravati2017nonreciprocal}, and temporal double-slit interference \cite{tirole2023double}. 
As an outstanding example of spatiotemporally modulated systems, a temporal waveguide which harnesses the total internal reflection of light at spatiotemporal boundaries and therefore confines pulses in between \cite{plansinis2015temporal, plansinis2016temporal}, provides a novel concept for guiding light. 
Up to date, previous researches on temporal waveguides focus on fundamental properties for realizing a single temporal waveguide \cite{plansinis2016temporal, plansinis2018cross}, while interactions between multiple temporal waveguides remain unexplored.

In this paper, we derive the fundamental formula for modeling interactions between two temporal waveguides, i.e., the spatiotemporal coupled-mode theory. 
Systematic parameters which determine temporal coupling coefficients are given, and hence our theory introduces a basic framework for studying the problem of coupled temporal waveguides. 
To showcase the capability of our formalism, we explore a temporal Mach-Zehnder interferometer (MZI) and further propose a design of a universal multiport interferometer \cite{reck1994experimental, clements2016optimal, pai2019matrix, bogaerts2020programmable} in the time domain for optical pulses. 
Such a universal multiport interferometer enables an arbitrary temporal transformation for sequential-propagating pulses within one spatial waveguide under the spatiotemporal modulation, which could find potential applications 
in optical signal processing.
Our work hence provides a useful theoretical tool in the arising field of spatiotemporal metamaterials \cite{lee2020metamaterials, engheta2021metamaterials, galiffi2022photonics, yin2022floquet, yuan2022temporal, pacheco2022time} to develop new-generation active photonic devices.

\section*{Model}
\begin{figure}[htbp]
\centering
\includegraphics[width=0.5\textwidth ]{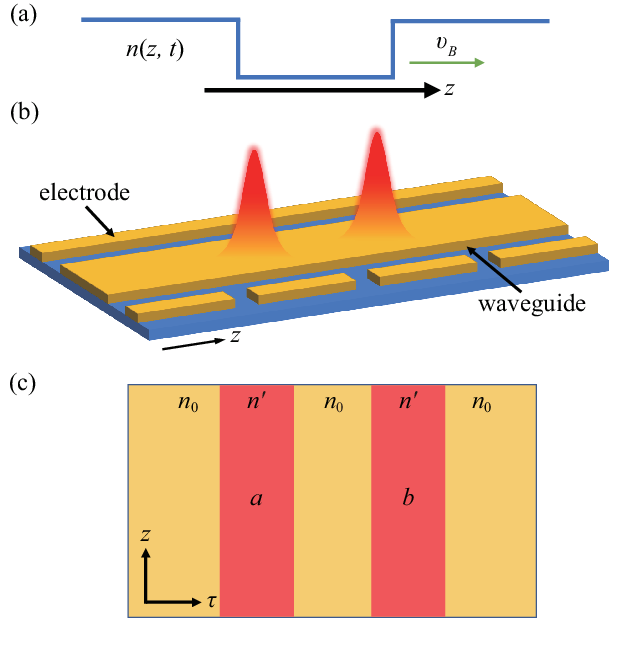}
\caption{(a) Schematic of the spatiotemporal modulation of the refractive index in a waveguide to achieve a temporal waveguide. (b) Scheme of the interaction of temporal waveguides, which is based on a waveguide modulated by traveling-wave signals through electrode arrays. (c) Refraction index $n(z,\tau)$ of the temporal waveguide $a$ and $b$ in the retarded frame $(z, \tau )$. Here $n'=n_0+\Delta n$.}\label{figure.1}
\end{figure}

We now start to show how to model interactions between two temporal waveguides and derive the coupled-mode formula in the time domain. 
Before getting into details, we first review the model of a temporal waveguide which is achieved with pulse propagating in a spatiotemporally modulated waveguide as shown in FIG.~\ref{figure.1}(a). The modulation of the refractive index of the waveguide is chosen as $n(z,t)=n_0+\xi (z-v_B t)$, where $n_0$ is the background refractive index of the waveguide and $\xi (z-v_B t)$ denotes the spatiotemporal change of the refractive index \cite{plansinis2016temporal} with $z$ being the propagating direction, and $v_B$ being the moving speed of the modulation. 
We transfer the formalism in a retarded time frame by using the transformation $\tau =t-z/v_B$, where $t$ is the time in the laboratory frame. 
In the retarded time frame $(z, \tau )$, the change of the refractive index $\xi(z-v_B t)$ is transformed to a $z$ independent function $\xi(\tau)=\xi_0+\Delta \xi(\tau)$. 
To achieve a temporal waveguide, $\xi(\tau)$ can be chosen as
\begin{eqnarray}\label{2}
    \xi(\tau) = \left\{ {\begin{aligned}
        &{{\Delta n \hspace{1em} |\tau-\tau_c|\leqslant T_w/2,}}\\
        &{{0 \hspace{2em} |\tau-\tau_c|>T_w/2,}}
        \end{aligned}} \right.
\end{eqnarray}
where $\Delta n$ is the modulation amplitude, and $T_w$ is the modulation time width centered at $\tau_c$ in the time retarded frame. As an analog to the conventional spatial waveguide, one can consider $|\tau-\tau_c|\leqslant T_w/2$ as the core region of the temporal waveguide, and $|\tau-\tau_c|>T_w/2$ as the cladding region. 

We then consider a pulse having a central frequency $\omega_0 $ propagates along such modulated waveguide. One can treat such a problem by Taylor-expanding the dispersion relation of the waveguide as \cite{plansinis2015temporal, plansinis2016temporal}:
\begin{eqnarray}\label{3}
\beta (\omega )=\beta_0+\Delta \beta _1 (\omega -\omega _0)+\frac{\beta _2}{2} (\omega -\omega _0)^2+\beta _m (\tau),
\end{eqnarray}
where $\beta_0=n_0\omega_0 /c$, $\Delta \beta _1=\beta _1-1/v_B$ with $\beta_1=\partial \beta /\partial \omega |_{\omega_0} $ being the reciprocal of the group velocity at $\omega_0$, $\beta _2=\partial^2 \beta /\partial \omega^2 |_{\omega_0}$ being the corresponding group velocity dispersion, and  $\beta _m (\tau)=\beta  _0 \hspace{0.25em} \xi  (\tau )/n_0$  represents the change of the propagation constant due to spatiotemporal modulation.
By using Maxwell's equation and the dispersion relation in Eq. (\ref{3}), one obtains the resulting wave equation for describing the amplitude of propagating pulse $A(z,\tau)$ in the retarded frame \cite{agrawal2013nonlinear}:
\begin{eqnarray}\label{4}
    \frac{\partial A(z,\tau)}{\partial z}+\Delta \beta _1 \frac{\partial A(z,\tau)}{\partial \tau} +i\frac{\beta _2}{2} \frac{\partial^2 A(z,\tau)}{\partial \tau^2} = i \beta _m (\tau)A(z,\tau).
\end{eqnarray}
Following the treatment in Ref. \cite{plansinis2016temporal}, one can take the modal solution $A(z,\tau)$ as $A(z,\tau)=M(\tau )\text{exp}[i(Kz-\Omega \tau )]$, where $M(\tau)$ describes temporal shape of the mode, $K$ denotes the rate of the mode accumulating phase during propagation, and $\Omega=-\Delta \beta _1/\beta _2 $ is the frequency shift. 
So far, we briefly outline the formalism of the optical pulse propagating in a waveguide under the spatiotemporal modulation of the refractive index, i.e., the field traveling inside a temporal waveguide in the retarded frame, which has been utilized in many follow-up studies \cite{plansinis2016temporal, plansinis2018cross, gaafar2019front, cai2020reflection}.

Next we give the important formula of the temporal coupled-mode theory. We construct two temporal waveguides (labeled as $a$ and $b$), which is achieved by spatiotemporally modulating the refractive index of a spatial waveguide through the traveling-wave signal by segmented electrodes as shown in FIG.~\ref{figure.1}(b). In particular, the spatiotemporal change of the refractive index in the retarded time frame is taken as 
\begin{eqnarray}\label{aa}
\xi' (\tau)=\xi_0+\Delta \xi_a (\tau)+ \Delta \xi_b (\tau),
\end{eqnarray}
where $\Delta \xi_a(\tau)$ and $\Delta \xi_b (\tau)$ follow Eq. (\ref{2}) centered at $\tau_c^a$ and $\tau_c^b$ $(|\tau_c^b-\tau_c^a|>T_w)$ and therefore form two temporal waveguides $a$ and $b$, respectively.

We consider the field at the central frequency $\omega _0$ propagating in such spatiotemporally modulated waveguide and assume a solution as
\begin{eqnarray}\label{5}
A(z,\tau)=G_a(z)M_a(\tau)\text{exp}[i(K_az-\Omega \tau )]+G_b(z)M_b(\tau)\text{exp}[i(K_bz-\Omega \tau )],
\end{eqnarray}
where $G_a$ ($G_b$) represents the envelope amplitude of the pulse in the temporal waveguide $a$ ($b$), and the corresponding $M_i(\tau)$ and $K_i(\tau)$ $(i=a, b)$ satisfy \cite{plansinis2016temporal}, 
\begin{eqnarray}\label{6}
    \frac{\partial^2 M_i(\tau)}{\partial \tau^2}+\frac{2}{\beta 
     _2}[K_i+\frac{(\Delta  \beta _1)^2}{2\beta _2}-\beta 
      _{m}^i(\tau) ]M_i(\tau)=0,
\end{eqnarray}
where $\beta _{m}^i(\tau)=\beta _0 \hspace{0.25em} \xi_i (\tau)/n_0$.
Substituting Eq. (\ref{5}) into Eq. (\ref{4}), multiplying by $M_a^*(\tau)$ or $M_b^*(\tau)$, and then integrating over $\tau$, we obtain the temporal coupled-mode equations:
\begin{subequations}\label{7}
\begin{align}
\frac{\partial G_a(z)}{\partial z} &=i\frac{\kappa _{ab}-C_{ab}\gamma _{b}}{1-C_{ab}C_{ba}} \text{exp}[i(K_b-K_a)z]G_b(z)+i\frac{\gamma  _{a}-C_{ab}\kappa_{ba}}{1-C_{ab}C_{ba}} G_a(z), \tag{7a}\label{8a}\\
\frac{\partial G_b(z)}{\partial z} &=i\frac{\kappa _{ba}-C_{ba}\gamma  _{a}}{1-C_{ba}C_{ab}} \text{exp}[-i(K_b-K_a)z]G_a(z)+i\frac{\gamma  _{b}-C_{ba}\kappa _{ab}}{1-C_{ba}C_{ab}} G_b(z), \tag{7b}\label{8b}
\end{align}
\end{subequations}
where $C_{ij}$, $\kappa _{ij}$, and $\gamma _i$ $(i\neq j)$ are expressed as,
\begin{subequations}\label{8}
\begin{align}
        C_{ij}&=\int M_i^*(\tau)M_j(\tau)d \tau /\int M_i^* 
         (\tau)M_i(\tau)d \tau, \tag{8a}\label{9a}\\
         \kappa _{ij}&=\int M_i^*(\tau) \Delta \beta 
          _{m}^i(\tau) M_j(\tau)d \tau/\int M_i^* 
          (\tau)M_i(\tau)d \tau, \tag{8b}\label{9b}\\
          \gamma _{i}&=\int M_i^*(\tau) \Delta \beta 
          _{m}^j(\tau) M_i(\tau)d \tau/\int M_i^* 
          (\tau)M_i(\tau)d \tau \tag{8c}\label{9c},
\end{align}
\end{subequations}
and $\Delta \beta _{m}^i(\tau)=\beta _0 \hspace{0.25em} \Delta \xi_i (\tau)/n_0$.
If we further assume the two temporal waveguides are identical in temporal shapes and $\int M_i^*(\tau)M_j(\tau)d \tau \ll \int M_i^*(\tau)M_i(\tau)d \tau$, we get $K_a=K_b$ and $C_{ab}\approx C_{ba}\approx 0$. Eqs. (\ref{8a})-(\ref{8b}) can then  be simplified as,
\begin{subequations}\label{10}
\begin{align}
    \frac{\partial G_a(z)}{\partial z} \approx i \kappa _{ab}
     G_b(z)+i \gamma _{a} G_a(z), \tag{9a}\label{10a}\\
    \frac{\partial G_b(z)}{\partial z} \approx i \kappa _{ba}
     G_a(z)+i \gamma _{b} G_b(z), \tag{9b}\label{10b}
\end{align}
\end{subequations}
Here, $\kappa _{ij}$ is the temporal coupling coefficient, and $\gamma _{i}$ is the shift due to the presence of the other temporal waveguide. Note that we consider a symmetry case here, so we can take $\kappa_{i, j}=\kappa_{j, i}^*$ and $\gamma_a=\gamma_b$.

\begin{figure}[htbp]
    \centering
    \includegraphics[width=0.6\textwidth ]{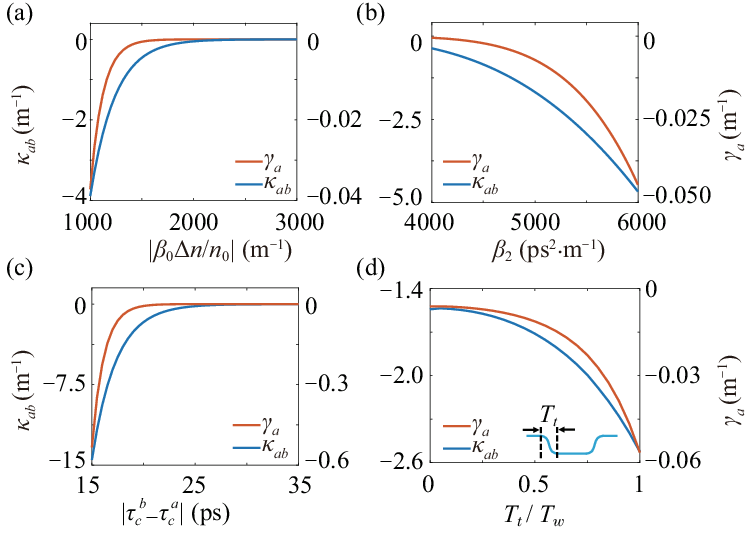}
    \caption{$\gamma _{a}$ and $\kappa _{ab}$ versus (a) $|\beta _0 \Delta n/n_0|$, (b) $\beta _2$, (c) time spacing between 
two temporal waveguides, and (d) turn-on/off time width $T_t$.}\label{figure.2}
\end{figure}

To give an illustrative picture on how the systematic parameters of this spatiotemporally modulated waveguide determine the coefficients in the temporal coupled-mode equations (\ref{10}), we give an example with experimentally-feasible parameters. 
We choose $\beta _0\Delta n/n_0=-1200~\text{m}^{-1}$, $\Delta \beta_1 =0$, $\beta _2=5000~\text{ps}^2\cdot \text{m}^{-1}$, which are standard parameters for an optical waveguide with the modulation strength $\Delta n/n_0 \sim 10^{-4}$ \cite{luennemann2003electrooptic,  li2020lithium, zhang2021integrated}. The spatiotemporal modulation shape can take $T_w=10$ ps, and $|\tau_c^b- \tau_c^a|= $20 ps. 
We now tune one of these parameters and fix others to investigate how $\gamma _{a}$ and $\kappa _{ab}$ are affected. 
Only the coupling between fundamental modes is considered (for the definition of the fundamental mode in a temporal waveguide, one can refer to \cite{plansinis2016temporal}). We first change $|\beta _0 \Delta n/n_0|$ as shown in FIG.~\ref{figure.2}(a). 
The role of $|\beta _0 \Delta n/n_0|$ in temporal waveguides is similar to the index contrast between the core region and cladding region in spatial waveguides. When $|\beta _0 \Delta n/n_0|$ becomes larger, both $\gamma _{a}$ and $\kappa _{ab}$ becomes weaker. Next we vary the dispersion parameter $\beta _2$, which describes the ability of light to spread out of the core region in the temporal waveguide. 
As a result, a larger $\beta _2$ results in larger $\gamma _{a}$ and $\kappa _{ab}$ as shown in FIG.~\ref{figure.2}(b). 
Fig.~\ref{figure.2}(c) shows changes of $\gamma_a$ and $\kappa_{ab}$ versus the time spacing between two temporal waveguides $|\tau_c^b-\tau_c^a|$, and one can see both coefficients decreases when the time spacing increases as two temporal waveguides fall apart in the time domain. 
In all calculations above, we consider the ideal modulations, i.e., the change of $\Delta \beta_m^i(\tau)$ is abrupt. In reality, the turn-on/off of modulations is not instantaneous. To reflect this feature, we consider the form of $\Delta \beta_m^i (\tau)$ as
\begin{eqnarray}\label{11}
    \Delta \beta_m^i (\tau) = \left\{ {\begin{aligned}
        &{{\beta_0  \Delta n /n_0 \hspace{11.8em} 
         |\tau-\tau_c^i| <\frac{1}{2} (T_w-T_t),}}\\
        &{{\beta _0 \Delta n/n_0 \hspace{0.2em} \text{cos}[|\tau-\tau_c^i |-\frac{1}{2} (T_w- 
         T_t)]} 
         \hspace{0.5em}
         \frac{1}{2} (T_w-T_t)\leqslant |\tau-\tau_c^i|<\frac{1}{2} (T_w+T_t),}\\
        &{{0 \hspace{15.3em} \text{others},}}
        \end{aligned}} \right.
\end{eqnarray}
where $T_t$ denotes the turn-of/off time width between the temporal core and the cladding regions [see the subfigure in FIG.~\ref{figure.2}(d)]. One can find that in FIG.~\ref{figure.2}(d), when $T_t/T_w$ becomes larger, $\gamma _{a}$ and $\kappa _{ab}$ are increasing, indicating that the confinement of the pulse is weaker. 
Nevertheless, when $T_t \sim 10\% \cdot T_w$, both coefficients do not change much compared to those when $T_t=0$, i.e, the ideal modulation case. Therefore, in the following, we still simulate models under the ideal modulation case.
\section*{Results}
\begin{figure}[htbp]
    \centering
    \includegraphics[width=0.9\textwidth ]{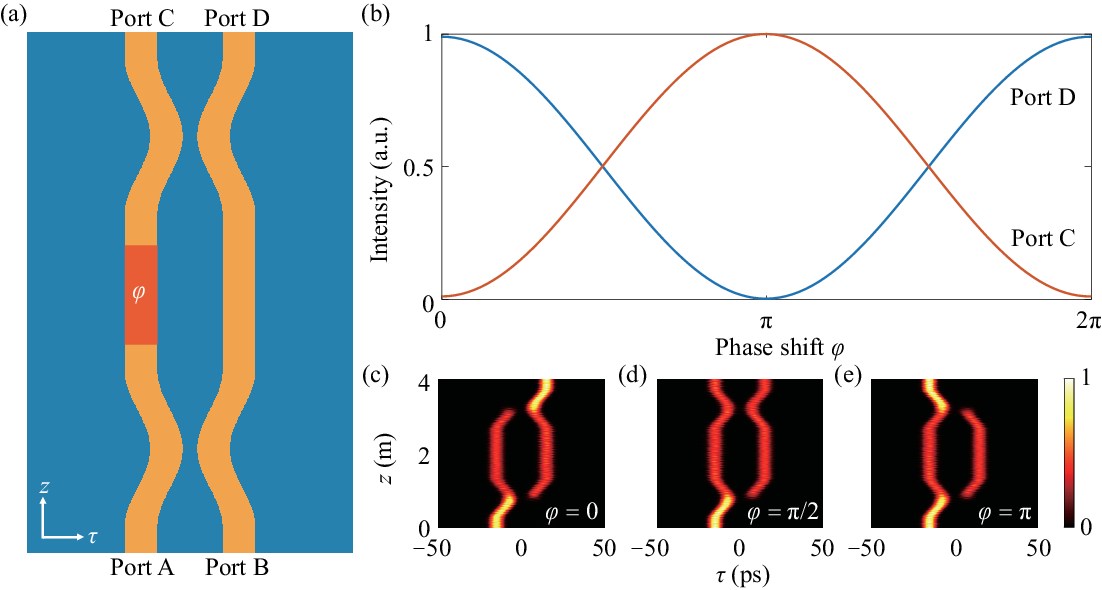}
    \caption{(a) Schematic of the temporal MZI. (b) Output intensities at Port C and Port D versus the phase shift $\delta $ in simulations. (c)-(e) The intensity distributions of the fields for the choice of (c) $\delta =0$, (d) $\delta =\pi /2$, and (e) $\delta =\pi$.}\label{figure.3}
\end{figure}

So far, we derive the spatiotemporal coupled-mode equations for modeling interactions between two temporal waveguides and investigate how systematic parameters of the system affect temporal coupling coefficients. 
In the following, we use these equations and demonstrate a temporal MZI with parameters $\beta  _0\Delta n/n_0=-1200~\text{m}^{-1}$, $\Delta \beta_1 =0$, $\beta _2=5000~\text{ps}^2\cdot ~\text{m}^{-1}$, and $T_w=10$ ps. 
The scheme of such temporal MZI is depicted in FIG.~\ref{figure.3}(a), with values of $\beta_m (z, \tau )=1200~\text{m}^{-1}$ in the cyan regime and $=0~\text{m}^{-1}$ in the orange regime of the retarded frame $(\tau,z)$. 
We aim to design the system with the functionality composed of two $50:50$ couplers and a phase shifter at one of the waveguides in the time domain. 
The parameters are selected to guarantee that $\kappa_{ij}\approx 0$ in the straight temporal waveguide region in FIG.~\ref{figure.3}(a) while $\kappa_{ij}\gg 0$ in the curved temporal waveguide region. 
The additional phase shift $\varphi $ is realized by an additional change of the refractive index $\delta n$ for the length $0.4~\text{m}$ corresponding to the red region in FIG.~\ref{figure.3}(a), resulting in a small change of $\beta_m(z,\tau)$, i.e., the choice of $\delta n/\Delta n \in [0,1.375\times 10^{-2}]$ gives effective phase shift $\phi \in [0,2\pi]$. 
We assume the input at Port A (Port B) as A (B), and the output at Port C (D) as C (D). The relation  for a temporal MZI is described by
\begin{eqnarray}\label{12}
\begin{pmatrix}
    C\\
    D
\end{pmatrix}
=
\begin{pmatrix}
    \text{sin}\frac{\delta }{2} & \text{cos}\frac{\delta }{2}\\
    \text{cos}\frac{\delta }{2} & -\text{sin}\frac{\delta }{2}\\
\end{pmatrix}
\begin{pmatrix}
    A\\
    B
\end{pmatrix},
\end{eqnarray}
which can be verified by results in our simulation given in FIG.~\ref{figure.3}(b). In particular, the output at Port C increases as $\varphi  $ increases and reaches its maximum at $\varphi  =\pi$ ($\delta n/\Delta n=6.875\times 10^{-3}$), while it further decreases and reaches its minimum at $\varphi  =2\pi$ ($\delta n/\Delta n=1.375\times 10^{-2}$). The output at Port D just behaves exactly in the opposite way.
In addition, three specific cases of $\varphi $ are taken and the intensity distributions of the field are plotted in FIGs.~\ref{figure.3}(c)-(e). 
In FIG.~\ref{figure.3}(c), we inject the pulse at Port A, and the pulse gradually switches to the other temporal waveguide during propagation. 
Such a phenomenon corresponds to a pulse traveling in a spatial waveguide and gradually converting to the other pulse in front of it spatially in the laboratory frame. 
In Fig.~\ref{figure.3}(d), the pulse gradually splits into two pulses which correspond to different spatial locations in one spatial waveguide, while in Fig.~\ref{figure.3}(e), the pulse temporally splits into two pulses then they converge to the original one eventually.

\begin{figure}[htbp]
    \centering
    \includegraphics[width=0.9\textwidth ]{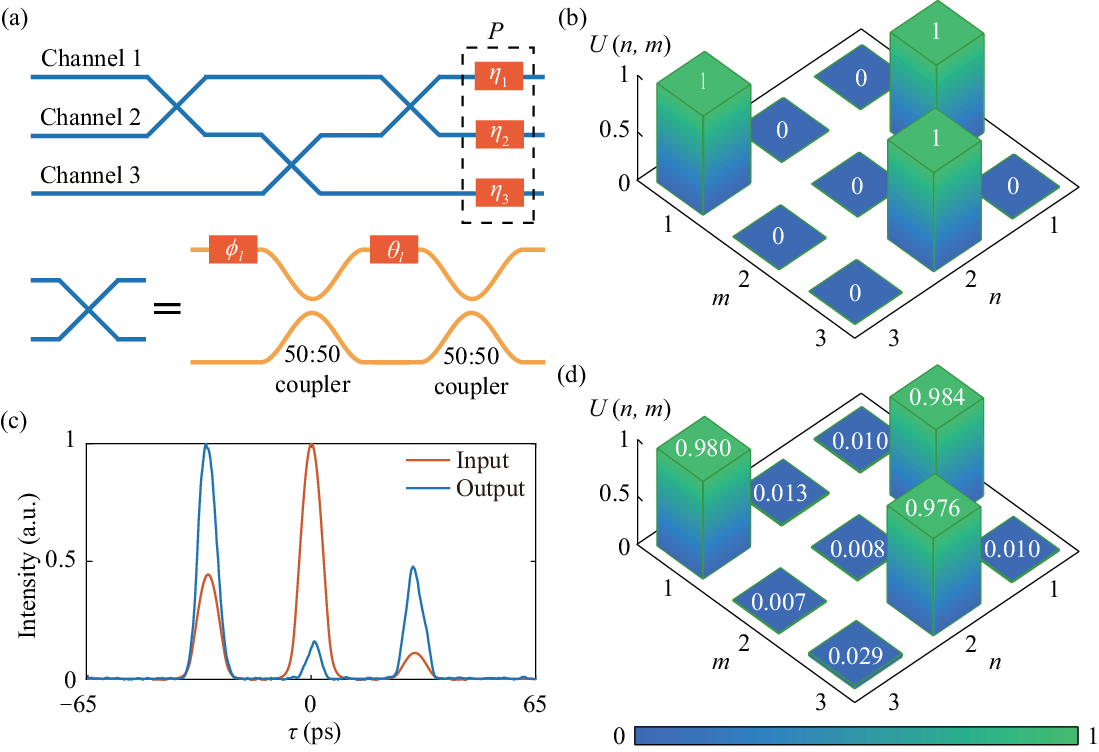}
    \caption{(a) Schematic of a three-channel temporal universal multiport interferometer. (b) The expected 
     transform matrix $U(n,m)$.
     (c) Normalized intensities at the input ports and output ports. (d) The reconstructed transform matrix $U(n,m)$ 
     from simulation results.}\label{figure.4}
\end{figure}

The proposed temporal MZI could have potential applications in optical signal processing and optical communication. Here, we showcase it being a component of a temporal universal multiport interferometer. The parameters are the same as those used in the above example of MZI. An arbitrary unitary transformation $U$ performed by a temporal universal multiport interferometer with $N$ channels shown in FIG.~\ref{figure.4}(a) can be decomposed in the following form:
\begin{eqnarray}\label{13}
    U=P \left(\prod_{(m,n) \in S} \hspace{-0.25em} T_{m,n,l}\right).
\end{eqnarray}
Here the production follows an ordered sequence ($S$) of two-channel transformations \cite{clements2016optimal}, and
\begin{eqnarray}\label{14}
    T_{m,n,l}(\theta_l ,\phi_l  )=
    \begin{pmatrix}
        1 & 0 & \hspace{1em} & \cdots & \cdots & \hspace{1em} & 
         \hspace{1em} & 0\\
        0 & 1 & \hspace{1em} & \hspace{1em} & \hspace{1em} & 
         \hspace{1em} & \hspace{1em} & \hspace{1em} \\
         \vdots & \hspace{1em} & \ddots \hspace{1em} 
         &\hspace{1em} & \hspace{1em} & \hspace{1em} 
        &\hspace{1em} & \vdots \\
        \hspace{1em} & \hspace{1em} & \hspace{1em} & e^{i\phi_l  } 
         \text{sin}\frac{\theta_l }{2} & \text{cos}\frac{\theta_l } 
          {2} &\hspace{1em} &\hspace{1em}\\
          \hspace{1em} & \hspace{1em} & \hspace{1em} &e^{i\phi_l  
           } \text{cos}\frac{\theta_l }{2} & - 
            \text{sin}\frac{\theta_l } {2} \hspace{1em} 
             &\hspace{1em} &\hspace{1em} & \hspace{1em} \\
             \vdots & \hspace{1em} &\hspace{1em} &\hspace{1em} 
              &\hspace{1em} & \ddots & \hspace{1em} & 
               \vdots\\
               \hspace{1em} &\hspace{1em} &\hspace{1em} 
                &\hspace{1em} &\hspace{1em} &\hspace{1em} & 1 & 
                 0\\
                0 &\hspace{1em} &\hspace{1em} &\cdots &\cdots 
                 &\hspace{1em} & 0 &1
    \end{pmatrix},
\end{eqnarray}
is the $l$-th transformation in such sequence between two channels $m$ and $n$ ($m=n-1$), which is realized by a modified MZI with an additional phase shift $\phi_l$ and splitting parameter $\theta_l$ between channels $m$ and $n$ in the time domain as shown in FIG.~\ref{figure.4}(a). 
$P$ is a diagonal matrix with complex elements whose modulus are equal to one, corresponding to a phase shift $\eta _m$ for channel $m$. We perform a three-channel temporal transformation for the demonstration in principle. $U(n,m)$ is designed as a circulation operator on the three temporal channels as shown in FIG.~\ref{figure.4}(b). 
Three pulses with different amplitudes  (namely in normalized intensities as $1$, $4/9$, and $1/9$, respectively shown in Fig. 4(c)) are injected into the input ports. In FIG.~\ref{figure.4}(c) the pulses take the desired circulation from one temporal waveguide to the other. This result corresponds to sequential-propagating pulses switching their position during propagation in the spatial waveguide in the laboratory frame. In addition, we reconstruct $U(n,m)$ based on the simulation result as shown in FIG.~\ref{figure.4}(d), which is closely matched with the desired one in FIG.~\ref{figure.4}(b).

\section*{Discussion}
We finally make a discussion on the possibility of realizing the proposal in experiments. The parameters in the simulation are achievable with the state-of-art technology in photonics \cite{wang2018integrated, boynton2020heterogeneously, liu2021sub}. 
For a pulse with the central wavelength $\sim 1000 ~\text{nm}$, the corresponding coefficients give $\Delta n\thicksim  10^{-4}$ and $\beta_2 \sim 10^3~\text{ps}^2\cdot ~\text{m}^{-1}$, which have been demonstrated in experiments \cite{luennemann2003electrooptic,  li2020lithium, zhang2021integrated, kaushalram2019tunable}. By properly engineering the waveguide structure, one can further enlarge the group velocity dispersion $\beta _2$ \cite{zhang2009highly}. 

In summary, we build a formalism of the temporal coupled-mode equations to study interactions between temporal waveguides in a system where pulses propagate in a spatiotemporally modulated waveguide, and show how systematic parameters of the modulated system determine the temporal coupling coefficients in the theory. 
A temporal MZI is studied and further a temporal universal multiport interferometer in the time domain is proposed, which enables an arbitrary unitary transformation for sequential-propagating pulses. 
Our work provides a fundamental method, which is useful for optical signal processing in time domain. 
In particular, compared with the conventional methods with coupled waveguides in the spatial dimension \cite{kawanishi1998ultrahigh, hamilton2002100, miller2013self}, the generalization of the coupled temporal waveguides does not require the addition of devices in the space, and the temporal transformation can be performed in only one spatiotemporally modulated waveguide, which greatly reduces the spatial complexity and insertion loss form the connection between multiple devices. 
Moreover, such transformation is realized by the spatiotemporal modulation in an active way, which provides more flexibility in manipulating pulses. 
The temporal coupled-mode theory in Eqs. (\ref{7})-(\ref{10}) can be further utilized to model two temporal waveguides structure with different systematic parameters and/or modulations that hold complex $\Delta n$. 
In addition, the proposed scheme is also compatible with previous works for controlling a single pulse in one temporal waveguide to achieve pulse compression \cite{ryan1995pulse, wagner2004self, chamanara2019linear}, fast and slow light \cite{kolchin2008electro, thevenaz2008slow, boyd2009slow, li2015tunable}, and so on \cite{li2022single}, hence it can trigger further studies on not only pulse transformation but multifunctional control for pulse propagation with multiple coupled pulse channels in the time domain, which further offers a wealth of opportunities in optical signal processing.

\vspace{1cm}

\textbf{Acknowledgements}
The research was supported by National Natural Science Foundation
of China (12122407, 11974245, and 12192252), National Key Research and Development Program of China (No. 2021YFA1400900). L.Y. thanks the sponsorship from Yangyang Development Fund and the support from the Program
for Professor of Special Appointment (Eastern Scholar) at
Shanghai Institutions of Higher Learning.


\bibliographystyle{apsrev}

\bibliography{temporal_waveguide}

\end{document}